\documentclass{elsart}
\usepackage{epsfig}
\journal{Solid State Communication}
\begin{document}
\begin{frontmatter}
\title{Absence of Metal-Insulator-Transition and Coherent
Interlayer Transport in oriented graphite in parallel magnetic fields}
\date{\today}
\author[leipzig]{H. Kempa}
\author[leipzig]{H. C. Semmelhack}
\author[leipzig]{P. Esquinazi}
\author[campinas]{Y. Kopelevich}
 \address[leipzig]{Department of Superconductivity
and Magnetism, Institute for Experimental Physics II, University of
Leipzig, Linn{\'e}str. 5, D-04103 Leipzig, Germany}
\address[campinas]{Instituto de F{\'i}sica, Universidade Estadual de
Campinas, Unicamp 13083-970, Campinas, S\~{a}o Paulo, Brasil}

\begin{abstract}
Measurements of the magnetoresistivity of graphite with
a high degree of control of the angle between the sample and
magnetic field indicate that the
metal-insulator transition (MIT), shown to be induced by a magnetic field
applied perpendicular to the layers, does not appear in parallel
field orientation. Furthermore, we show that
interlayer transport is coherent in less ordered samples and high
magnetic fields, whereas appears to be incoherent in
less disordered samples. Our results demonstrate the two-dimensionality
of the electron system in ideal graphite samples.
\end{abstract}
\end{frontmatter}


Recent experimental and theoretical studies of graphite have
renewed the interest in this system
\cite{kopeA,kopeB,KempaA,Sercheli,KempaB,gonA,gonB,khvA,khvB,Gorbar}.
Experimental results show that, contrary  to the common belief,
the transport and magnetic properties of graphite cannot be accounted for by
semiclassical models.
Experiment and theory raise a number of
questions concerning the coupling between the graphite layers.
The understanding of the transport properties is of primary interest and can
provide a fundamental contribution to the physics of two-dimensional (2D) systems in general.
In this letter we deal with two important open
questions:

1. A MIT appears both in the in-plane
\cite{kopeA,KempaA,Sercheli} and out-of-plane \cite{KempaB}
resistivity induced by a magnetic field $B$ applied perpendicular to
the graphite layers, i.e. $B || c$-axis.
Based on  magnetization data \cite{kopeB} and the
found 2D scaling  similar to that in the MIT
of  Si-MOSFETs (metal-oxide-semiconductor field-effect-transitor) \cite{Phillips},
Mo-Ge  \cite{mason} and Bi-films \cite{marko},
the MIT in graphite has
 been interpreted in terms of superconducting fluctuations
\cite{kopeA,KempaA}.
It has been shown
that the density of states at the Fermi level is enhanced through
topological disorder, thus leading to the possible occurrence of
localized ferromagnetism and superconductivity \cite{gonB}.

Other interpretation of the MIT in graphite uses the idea that a single graphite
layer is the physical realization of the relativistic theory
of (2+1)-dimensional Dirac fermions due to the linear dispersion
in the spectrum of quasiparticle excitations in the vicinity of
the corners of the Brillouin zone \cite{Semenoff}.
Taking this into account and the large Coulomb coupling constant for graphite
\cite{KempaB,khvA},
a magnetic catalysis (MC) has been proposed
\cite{khvB}.
This original explanation assumes
that a magnetic field perpendicular to the graphite layers breaks the chiral symmetry
and opens a gap in the spectrum of the quasiparticles at the corners of
the Brillouin zone. This effect is interpreted as the enhancement of the fermion dynamical
mass through electron-hole pairing,  i.e. a transition to an excitonic insulating state.

Which role does the field direction play? The experimental evidence
indicates that the MIT in Si-MOSFETs occurs independently of the
orientation of the magnetic field and, therefore, has been considered to
be driven solely by spin-dependent effects \cite{Kravchenko}. On the other
hand, the MC in graphite would be possible only for the case $B || c$,
i.e. the transition should be absent in the parallel case \cite{khvA}. To
our knowledge there is no experimental proof published in the literature
that the MIT is absent in oriented graphite samples for fields applied
parallel to the graphite planes. A clear experimental evidence for the
absence or not of the MIT would give an important hint to search for its
origin. This is one of the tasks of our experimental work.

2. In a recent work we have shown that the metalliclike behavior
of  the out-of-plane $c-$axis resistivity $\rho_c$ is directly correlated to
that of the in-plane resistivity \cite{KempaB} and that the intrinsic $\rho_c(T)$
of an ideal graphite sample would not be metalliclike.
These results cast
doubts about the   interlayer
transfer integral value of $\sim 0.3$~eV used  all over the literature \cite{Kelly,gonA}
in which neither the electron-electron interaction nor
charge fluctuations were taken into account \cite{Vozmediano}. In
principle, for such a large interlayer transfer one would expect
coherent transport for the interlayer magnetoresistance at low
temperatures. In this work we provide an answer to the question
whether the $c-$axis transport in graphite is or not coherent and
whether this (in)coherent transport is influenced by sample inhomogeneities.

\begin{figure}
\begin{center}
\epsfig{file=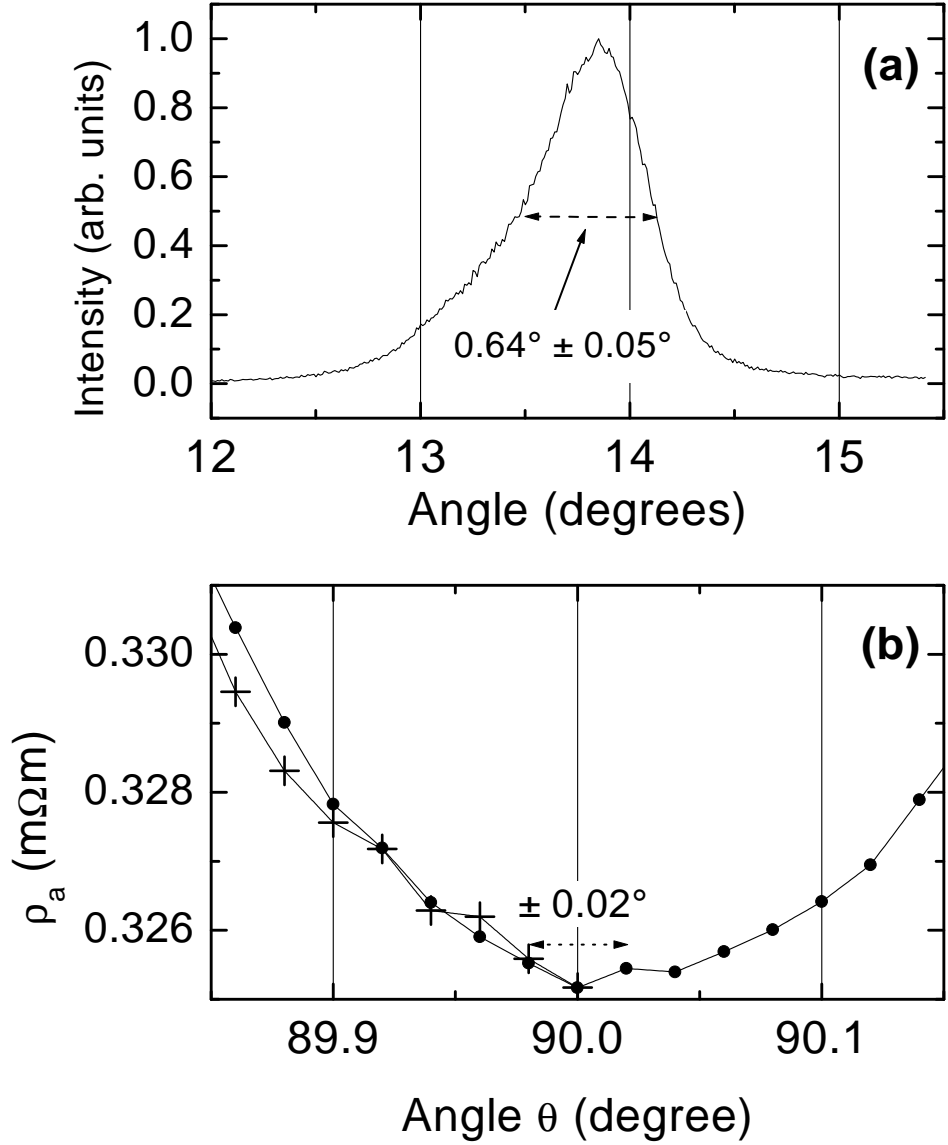,width=\columnwidth}
\end{center}
\vspace{-2.5cm}
\caption[*]{(a) X-ray rocking curve of the sample HOPG-3.
(b) Angle dependence of the in-plane resistivity ($\bullet$) for the same
sample in a magnetic field of 9~T at 2 K.
The field parallel to the sample at $90^{\circ}$ is defined at the
resistivity minimum. A second run ($+$) is stopped at
$90^{\circ}$.} \label{Align}
\end{figure}

 We have measured the
following samples:   a highly oriented pyrolytic graphite (HOPG) manufactured at the Research
Institute "Graphite" in Moscow and denoted as HOPG-3 \cite{KempaA}.
The X-ray rocking curve for this sample, see Fig. \ref{Align}(a),
 exhibits a full width at half
maximum (FWHM) of  $(0.64\pm0.05)^{\circ}$. We take this value as a
measure for the misalignment of the graphite layers within
the sample with respect to each other. Further
two HOPG samples, one from Union Carbide Corp. with a
 FWHM of $0.24^{\circ}$ (HOPG-1) and the other from
Advanced Ceramics Corp. with a FWHM of
$0.40^{\circ}$ (HOPG-2), have been measured. The fourth sample was
a Kish graphite sample with
a FWHM of  $1.6^{\circ}$ and a in-plane resistivity at 2K and
zero field $\sim 100$ times {\em smaller} than for the HOPG samples.
AC resistivity
measurements were performed by a conventional four-probe method.
The samples were fixed in a rotating sample holder inside the bore of
a 9T superconducting solenoid. Temperature stability was better than 2~mK
in the whole temperature range.

The crucial point of the magnetoresistance measurements in
parallel field is the misalignment of the
sample with respect to the field. The ultimate limit for the
alignment of the sample would be that of the
graphite layers within the sample. To adjust the sample to the
magnetic field we measured the angle dependence of the in-plane
resistivity with the current always perpendicular to the
field. The result is shown in Fig. \ref{Align}(b) for the HOPG-3 sample. Here
$90^{\circ}$, i.e.  field parallel to the sample, is
defined at the minimum of the resistivity. The high angle
resolution, the  strong angle dependence of the
resistivity and the high sensitivity of its measurement, as well as
 the excelent  reproducibility of the absolute angle (of the
order of the angle resolution, see Fig.~\ref{Align}(b)) allow us to align the sample
parallel to the field with
an accuracy of $\pm0.02^{\circ}$, well below the
 FWHM of the rocking curve.
\begin{figure}
\vspace{-0.3cm}
\begin{center}
\epsfig{file=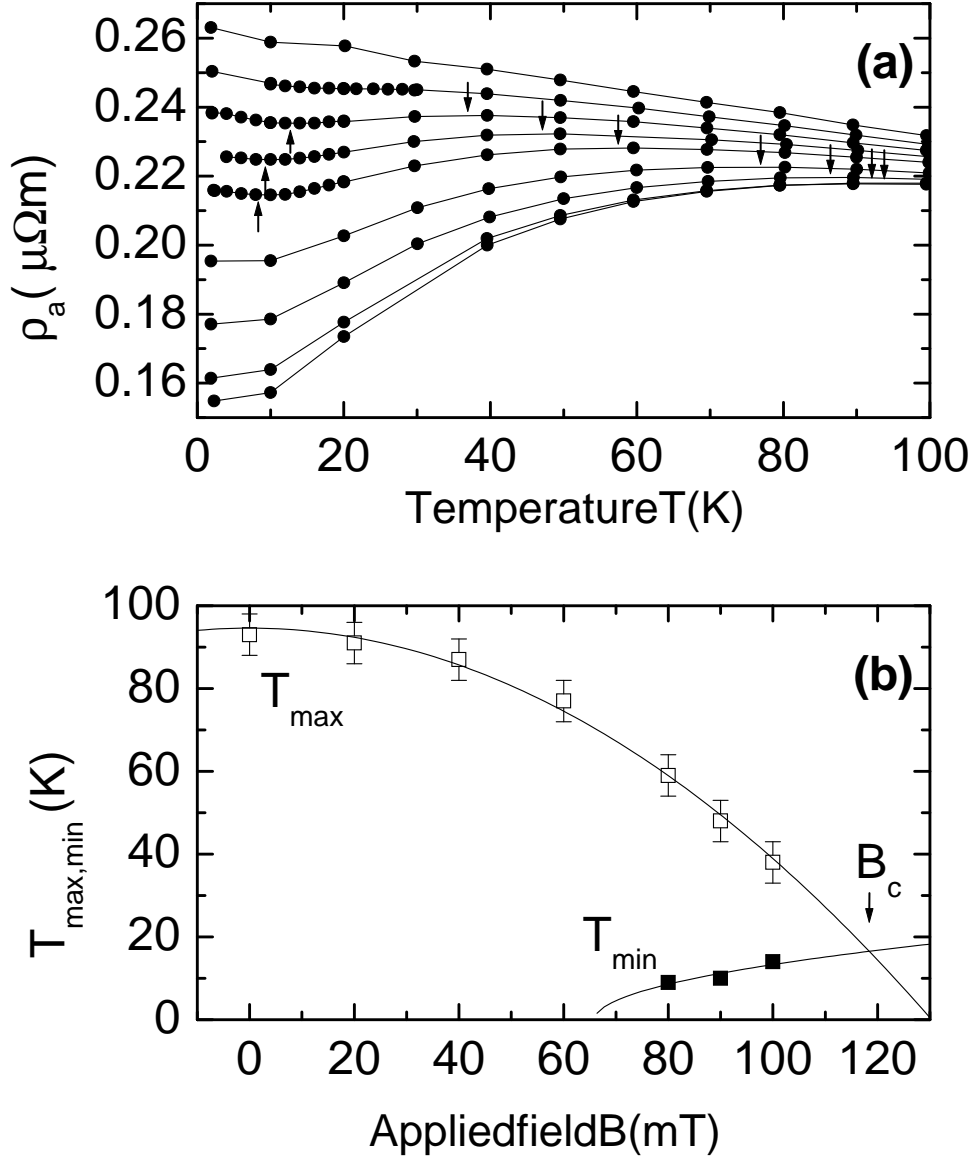,width=\columnwidth}
\end{center}
\vspace{-2.7cm}
\caption[*]{
(a) In-plane resistivity of the sample HOPG-3 as a function of
temperature for magnetic fields $B$ = 0, 0.02, 0.04, 0.06, 0.08,
0.09, 0.10, 0.11, 0.12 T (bottom to top) applied $|| c$. Upward (downward )
arrows mark the minima (maxima) of the resistivity.
(b) Temperatures of the maxima and minima of the resistivity as a
function of field together with fits for $T_{\rm max}$:  Eq.~(\protect\ref{Tmax}) $(-)$
with $B_1 = 0.13~$T;
$T_{\rm min}:$ dotted line
corresponds to Eq.~(\protect\ref{Tmin}) with $B_0 = 65.9~$mT.
The intersection
of the curves gives the critical field $B_{c}$.
} \label{BIIc}
\end{figure}

Figure \ref{BIIc}(a) shows the results  of the
in-plane resistivity with $B || c$. We note that the metalliclike phase, i.e. ${\rm d}\rho/{\rm d}T>0$,
is observed between a maximum  $T_{\rm max}$
and a minimum temperature $T_{\rm min}$ and for fields below
a critical field $B_{c}$ . The difference between
$T_{\rm max}$ and $T_{\rm min}$  decreases as the field approaches
$B_{c}$, i.e.  $T_{\rm max}(B_c) = T_{\rm min}(B_c)$, and the metalliclike
phase disappears.  Figure \ref{BIIc}(b) shows these data together with
fits of $T_{\rm max}(B)$ to the experimentally  found \cite{kopeA} relation
\begin{equation}
    T_{\rm max}(B) \propto \left (1- \frac{B^2}{B_{1}^2}\right ) \,,
\label{Tmax}
\end{equation}
 and of  $T_{\rm min}(B)$
to  the relation~\cite{Sercheli}
\begin{equation}
    T_{c}(B) \propto \sqrt{B-B_0}\,,
\label{Tmin}
\end{equation}
where $B_0$ and $B_1$ are free parameters, the first is refered as the offset field \cite{Gorbar}.
 The extrapolation of those relations serves
 to determine the critical field at their crossing which  for
the present case is $B_{c}^{\perp}=(0.12\pm0.01)$T.
\begin{figure}
\vspace{-0.3cm}
\begin{center}
\epsfig{file=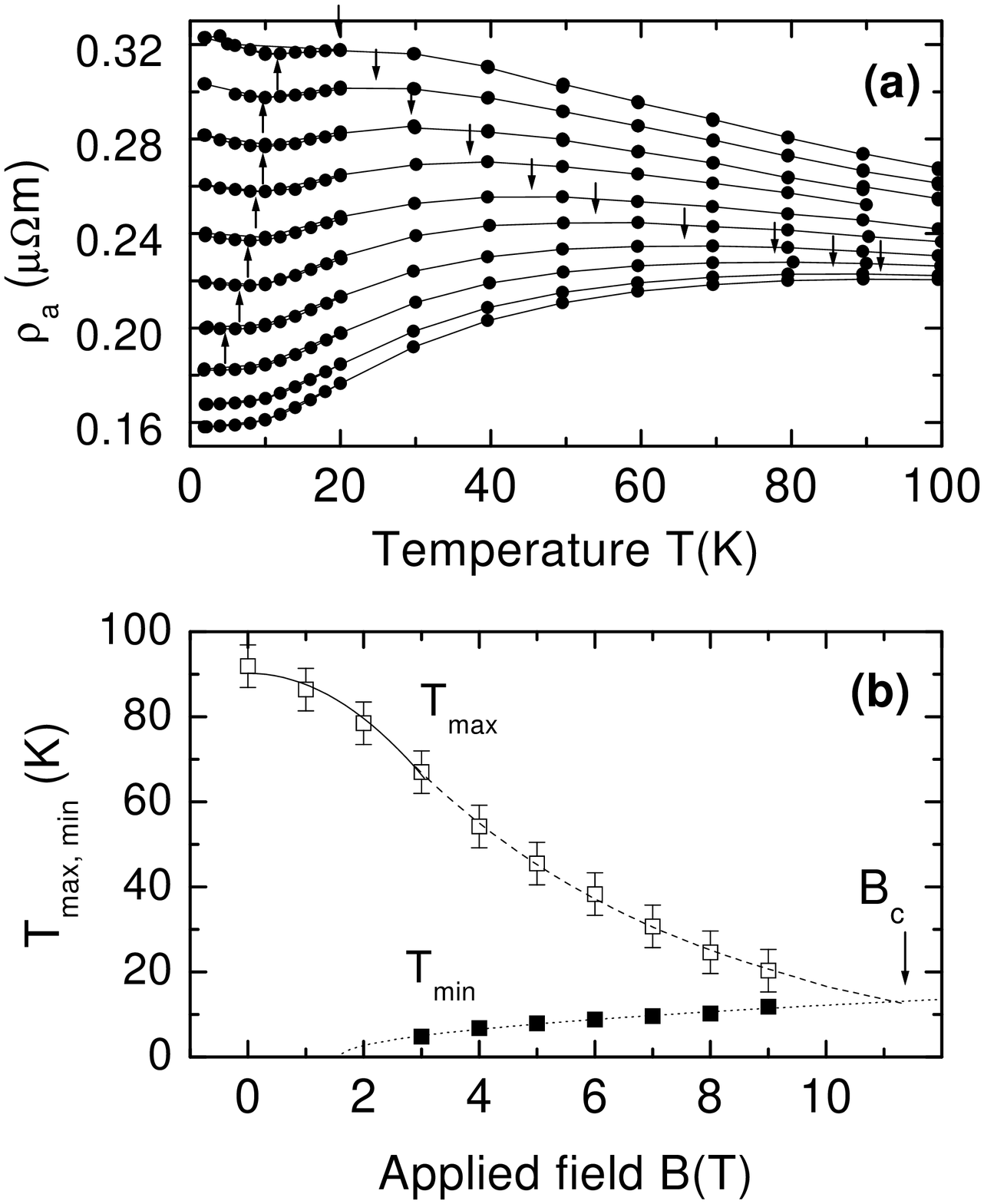,width=\columnwidth}
\end{center}
\vspace{-2.8cm}
\caption[*]{(a) In-plane resistivity of the sample HOPG-3 as a
function of temperature for magnetic fields $B = 0, 1\ldots 9~$T
(bottom to top) applied $\perp c$, aligned as in Fig. \ref{Align}(b).
(b) Temperatures of the maxima and minima of the in-plane resistivity as a
function of field together with fits for $T_{\rm max}$: Eq.~(\protect\ref{Tmax}) $(-)$
with $B_1=5.83~$T   and
dashed line  $\propto \exp(-B/B_2)$ with $B_2=5.11$~T; $T_{\rm min}:$ dotted line
corresponds to Eq.~(\protect\ref{Tmin})
 ($B_0 = 1.59~$T).} \label{BIIab}
\end{figure}

The results obtained for $B \perp c$  (aligned as shown in Fig. \ref{Align}(b)) are shown in Fig.~\ref{BIIab}(a)
and the corresponding $T_{\rm min}$ and $T_{\rm max}$
in Fig.~\ref{BIIab}(b). Whereas $T_{\rm min}(B)$
follows a similar relation as in the former case, $T_{\rm max}(B)$ is
reproduced by Eq.~(\ref{Tmax}) only in the low field range. At high fields
 the empirical relation $T_{\rm max}\propto
e^{-\frac{B}{B_{2}}}$ is found.
We note that the intersection of $T_{\rm max}(B)$
and $T_{\rm min}(B)$ is at $B_{c}^{\parallel}=(11.3\pm0.1)$T.
The dominant contribution to the misalignment is the
misalignment of the graphite layers with respect
to each other and is taken to be $(0.64\pm0.05)^{\circ}$ (see Fig.~1). The
perpendicular component of  $B_{c}^{\parallel}$ is then
$(0.13\pm0.01)$T. This value is within the error equal to
$B_{c}^{\perp}$. From this we conclude that the MIT is, if not
solely driven by the magnetic field perpendicular to the graphite
layers, by far dominated by it.

We would like to note two details of the field-driven transition shown in
Fig.~2(a) for sample HOPG-3, as well as in previous publications
\cite{KempaA,Sercheli,KempaB}. First, in the ``insulating" side of the
transition at $B \ge B_c$ we would expect that the resistivity $\rho
\rightarrow \infty$ for $T \rightarrow 0$. Instead we observe always a
saturation or, upon applied field, a weak logarithmic increase of the
resistivity decreasing temperature \cite{review}. The saturation for $T
\rightarrow 0$ in the insulating side of the transition has been reported
for various 2D systems as, for example, in Mo-Ge films \cite{mason},
GaAs/AlGaAs heterostructures \cite{prl1} and Josephson junction arrays
\cite{prl2}. There is no consent on the origin of this saturation.
Whatever the reason for it is, the main result of this part of the work,
i.e. that only the normal component of the applied field to the planes
drives the transition, remains untouched.

Second, one would tend to underestimate the real magnitude of the effect
driven by the applied field because of the relative small change of the
resistivity shown in Fig.~2(a). However, we stress that the relative
change in the resistivity with field depends on the sample
characteristics. Experimental data from different graphite samples show a
change at low temperatures between $\sim 20\%$ up to more than one order
of magnitude for a field of the order of 1~kOe
\cite{KempaA,Sercheli,KempaB,thesis}. Qualitatively speaking, the
transition is similar for all samples studied.

We discuss now shortly the origin of the MIT in graphite. First we note that
Eq.~(72) from Ref.~\cite{Gorbar}
$(T_{c}(B) \propto (1-({B_{0}^2}/{B^2}))\sqrt{B})$  fits the data for $T_{\rm min}(B)$
as well as Eq.~(\ref{Tmin}) with similar $B_0$. In Ref.~\cite{Gorbar} the authors
argue that the offset field $B_0 \simeq \pi n c / |e|$ ($n$ is the charge density)
 is model independent and is related to
the minimum field required to fill the lowest Landau level, necessary condition
to deblock the electrons for  pairing and to produce the excitonic gap. For the majority
carriers $n \sim 10^{11}$cm$^{-2}$ and
 $B_0 \sim 2~$T in clear disagreement with the experimental result  $\sim 0.06~$T.
But, if we take the minority carrier density  $n \sim  10^{9}$cm$^{-2}$
\cite{Kelly} we get roughly the measured $B_0$.  Nevertheless and
since MC is a 2D phenomenon, it is unclear
 whether this assumption is valid. We note that
recent results \cite{KempaB} including those from this work indicate that
for  HOPG samples, the coupling between planes is much weaker than
previously assumed, casting doubts about
the correctness of a 3D Fermi surface with two types of carriers
with different densities and effective
masses for ideal graphite \cite{Kelly}.

Returning to the superconducting scenario which is supported
by the magnetization data \cite{kopeB}, we may relate
 $T_{\rm max}(B)$  to the critical temperature of a system of
superconducting islands in a semiconducting
matrix \cite{kopeA}. In this case mesoscopic effects play a role.
Indeed, the observed
 behavior in the parallel case
 shows a change of curvature of  $T_{\rm max}(B)$ similar to that
found theoretically for disordered 2D superconductors \cite{spivak}.
We note also that an exponential decay with field for  $T_c(B)$ has been predicted
for 2D superconductors with weakly Josephson-coupled local superconducting islands
\cite{gali}. Because in the parallel case  the MIT occurs at large
fields, we are able to observe the anomalous decay of  $T_{\rm max}$
 in contrast to the transverse case, see Fig.~\ref{BIIc}(b).
On the other hand we may argue that
the reason for the difference in  the behavior of $T_{\rm max}(B)$ may be due
to its sensitivity to the intrinsic misalignment of the graphene planes
of the sample that shows a gauss-like distribution.

\begin{figure}
\begin{center}
\epsfig{file=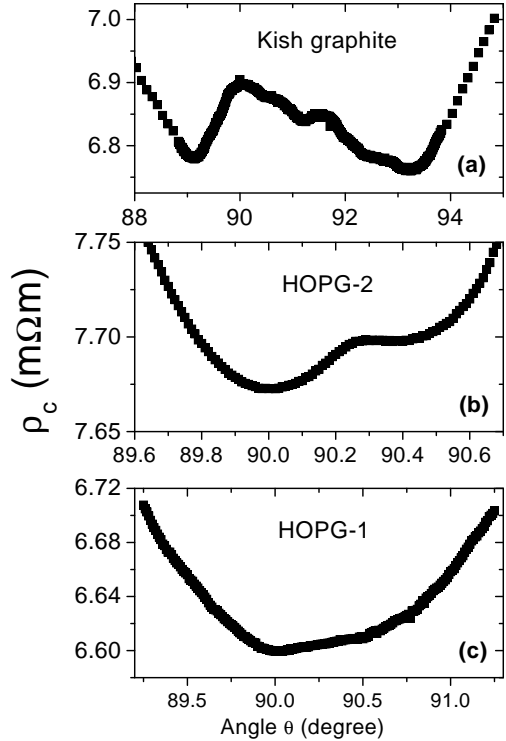,width=7cm}
\end{center}
\caption[*]{Angle dependence around $90^\circ$ of the out-of-plane resistivity of
the Kish graphite (a), HOPG-2 (b) and HOPG-1 (c) samples at $B = 9~$T and
at 2~K.} \label{Peak}
\end{figure}

We discuss now the interlayer transport. One possible way to test coherent
transport across the graphite layers is given by the measurement of a
maximum in the angle dependence of  $\rho_c(\theta)$ at magnetic fields
parallel to the layers \cite{Moses,Singleton,Wosnitza}. This peak should
be absent for incoherent interlayer transport but observed if the
inequality $\omega_c \tau  >1$ holds, where $\omega_c$ is the cyclotron
frequency and $\tau$ the relaxation time of the carriers. Coherent
transport means therefore that band states extend over many layers and a
3D Fermi surface can be defined. In the other case, incoherent transport
is diffusive and neither a 3D Fermi surface nor the Bloch-Boltzmann
transport theory is applicable \cite{Moses}. In order to check this and
the role played by disorder we have performed measurements of the
out-of-plane electrical resistivities with  high angle resolution.

 Figure \ref{Peak} shows the
results for three samples  at 9~T and 2~K.
We observe that a weak coherent peak in $\rho_c$ around the parallel orientation
($90^{\circ}$)  occurs and this is larger the larger the FWHM of the corresponding
rocking curve. The
asymmetry seen in the angle dependence is in part due to the small
experimental misalignment of the
surface of the sample and to the lack of crystal perfection. The coherent
peak decreases as expected with field.
Our results indicate that lattice defects not only affect the transport
as scattering centers but they contribute to enhance the coupling between the
layers giving rise to a 3D-like electronic spectrum and coherent transport.
The absence of coherent peak  in ideal samples may be related either to
incoherent transport or that $\omega_c \tau  < 1$ holds. Although the
validity of semiclassical criteria for incoherent transport is under discussion \cite{Singleton}
we use the Ioffe-Regel-Mott maximum
metallic resistivity $\rho_{\rm max}$ criterion to evaluate coherent transport \cite{xie}.
We obtain that only for the HOPG-1 and -2 samples $\rho_c > \rho_{\rm max}$ in the whole $T-$ and $B-$range,
in agreement with the absence of coherent peak.

From our results we draw the following conclusions:
 The MIT in HOPG is triggered only by  a magnetic field
perpendicular  to the graphite layers.  Therefore, it is unlikely that
spin effects play a significant
role in the MIT. The absence of  the MIT in the parallel field
orientation supports the theoretical approach of  Refs.~\cite{khvA,khvB,Gorbar}, but
there is apparently no quantitative agreement \cite{Gorbar}. On the other hand, the influence of
possible superconducting fluctuations on the MIT cannot be ignored.
 The transport perpendicular to the graphite layers in highly oriented and less disordered
samples appears to be incoherent, demonstrating the quasi-2D character of the
electron system of graphite. Samples defects lead to
a better coupling between the layers, a 3D-like behavior and coherent
interlayer transport, added to the possible local enhancement of
the density of states which may generate local superconductivity and ferromagnetism
\cite{gonB}.

\ack

This work is supported by the Deutsche Forschungsgemeinschaft under DFG ES
86/6-3. We acknowledge discussions with F. Guinea,  M.A.H. Vozmediano and
M. P. Lopez-Sancho. Y.K. was supported by CNPq and FAPESP. P.E.
acknowledges the hospitality of the Condensed Matter Physics Department
(C-III) of the Universidad Aut\'onoma de Madrid and the support given by
the Secretar\'ia de Estado de Educaci\'on y Universidades (grant
SAB2000-0139).

\end{document}